\documentclass[cernpreprint,texlive=2011,txfonts,UKenglish,texmf,paper=letter]{atlasdoc}
\usepackage[biblatex=false]{atlaspackage}
\usepackage{atlasphysics}

\usepackage{amssymb}
\usepackage{amsmath}

\usepackage{graphicx}
\usepackage{bm}

\usepackage{xspace}
\usepackage{twoopt}

\usepackage{rotating}

\usepackage{color}

\usepackage{cite}

\newcommand{\beq}{\begin{equation}}
\newcommand{\eeq}{\end{equation}}
\newcommand{\bea}{\begin{eqnarray}}
\newcommand{\eea}{\end{eqnarray}}

\def\powhegbox{{\sc Powheg Box}}
\def\pythia{{\sc Pythia}}
\def\madgraph{{\sc Madgraph}}

\def\bsp#1\esp{\begin{split}#1\end{split}}
\def\bpm{\begin{pmatrix}}
\def\epm{\end{pmatrix}}

\newcommand\cyreslimone{[-7.4, 5.7]$\times10^{-4}$}
\newcommand\cyreslimtwo{[3.8, 5.1]$\times10^{-3}$}

\newcommand\tcyreslim{[-1.8, 1.8]$\times10^{-3}$}

\newcommand\cgreslimone{[-5.8, -3.8]$\times10^{-4}$}
\newcommand\cgreslimtwo{[-0.7, 1.3]$\times10^{-4}$}

\newcommand\tcgreslim{[-2.4, 2.4]$\times10^{-4}$}

\newcommand\cHWreslim{[-8.6, 9.2]$\times10^{-2}$}
\newcommand\tcHWreslim{[-0.23, 0.23]}

\AtlasTitle{Constraints on non-Standard Model Higgs boson interactions in an effective Lagrangian using differential cross sections measured in the $H \to \gam\gam$ decay channel at $\sqrt{s} = 8$ TeV with the ATLAS detector}

\AtlasRefCode{HIGG-2015-02}

\date{\today}

\AtlasJournal{Physics Letters B}
\PreprintIdNumber{CERN-PH-EP-2015-182}

\AtlasAbstract{
The strength and tensor structure of the Higgs boson's interactions are investigated 
using an effective Lagrangian, which introduces additional 
CP-even and CP-odd interactions 
that lead to changes in the kinematic properties of the Higgs boson and associated jet spectra with respect to the Standard Model. The parameters of the effective Lagrangian are probed using a fit to five differential cross sections previously measured by the ATLAS experiment in the $H \to \gam\gam$ decay channel with an integrated luminosity of 20.3~fb$^{-1}$ at $\sqrt{s}=8$~TeV. In order to perform a simultaneous fit to the five distributions, the statistical correlations between them are determined by re-analysing the $H \to \gam\gam$ candidate events in the proton-proton collision data. No significant deviations from the Standard Model predictions are observed and limits on the effective Lagrangian parameters are derived. The statistical correlations are made publicly available to allow for future analysis of theories with non-Standard Model interactions.
}

\newcommand\widthfactor{0.7}

\begin{document}

\section{Introduction}
\label{sec:intro}

The discovery of a Higgs boson at the ATLAS and CMS experiments~\cite{Aad:2012tfa,Chatrchyan:2012ufa} offers a new opportunity to search for physics beyond the Standard Model (SM) by examining the strength and structure of the Higgs boson's interactions with other particles. Thus far, the interactions of the Higgs boson have been probed using the $\kappa$-framework~\cite{Heinemeyer:2013tqa}, in which the strength of a given coupling is allowed to vary from the SM prediction by a constant value. In this approach, the total rate of a given production and decay channel can differ from the SM prediction, but the kinematic properties of the Higgs boson in each decay channel are unchanged.

An alternative framework for probing physics beyond the SM is the effective field theory (EFT) approach~\cite{Buchmuller:1985jz, Hagiwara:1993qt,Giudice:2007fh,Grzadkowski:2010es,Contino:2013kra,Heinemeyer:2013tqa}, whereby the SM Lagrangian is augmented by additional operators of dimension six or higher. 
Some of these operators produce new tensor structures for the interactions between the Higgs boson and the SM particles, which can modify the shapes of the Higgs boson kinematic distributions as well as the associated jet spectra. The new interactions arise as the low-energy manifestation of new physics that exists at energy scales much larger than the partonic centre-of-mass energies being probed. 

In this Letter, the effects of operators that produce anomalous CP-even and CP-odd interactions between the Higgs boson and 
photons, gluons, $W$ bosons and $Z$ bosons are studied using an EFT-inspired effective Lagrangian. 
The analysis is performed using a simultaneous fit to five detector-corrected differential cross sections in the $H\to \gamma\gamma$ decay channel, which were previously published by the ATLAS Collaboration~\cite{Aad:2014lwa}. These are the differential cross sections as functions of the diphoton transverse momentum ($p_{\rm T}^{\gamma\gamma}$), the number of jets produced in association with the diphoton system ($N_{\rm jets}$), the leading-jet transverse momentum ($p_{\rm T}^{ j_1}$), and the invariant mass ($m_{jj}$) and difference in azimuthal angle ($\Delta\phi_{jj}$) of the leading and sub-leading jets in events containing two or more jets. 
 The inclusion of differential information significantly improves the sensitivity to operators that modify the Higgs boson's interactions with $W$ and $Z$ bosons. To perform a simultaneous analysis of these distributions, the statistical correlations between bins of different distributions need to be included in the fit procedure. These correlations are evaluated by analysing the $H\rightarrow \gamma \gamma$ candidate events in the data, and are published as part of this Letter to allow future studies of new physics that produces non-SM kinematic distributions for $H\to \gamma\gamma$.

\section{Higgs effective Lagrangian}\label{sec:eft}

The effective Lagrangian used in this analysis is presented in Ref.~\cite{Contino:2013kra}. In this model, the SM Lagrangian is augmented with the dimension six CP-even operators of the Strongly Interacting Light Higgs formulation~\cite{Giudice:2007fh} and corresponding CP-odd operators. 
The $H\rightarrow\gamma \gamma$ differential cross sections are mainly sensitive to the operators that affect the Higgs boson's interactions with gauge bosons and the relevant terms in the effective Lagrangian can be specified by 
\begin{eqnarray*}\label{eq:effT}
 {\cal L}_{\rm eff} =  & \! \! \! \! \bar{c}_{\gamma} {\cal O}_{\gamma} +  \bar{c}_{g} {\cal O}_{g} + \bar{c}_{HW} {\cal O}_{HW} + \bar{c}_{HB} {\cal O}_{HB} \\
     + & \! \! \!  \tilde{c}_{\gamma} \tilde{\cal O}_{\gamma} +  \tilde{c}_{g} \tilde{\cal O}_{g} + \tilde{c}_{HW} \tilde{\cal O}_{HW} + \tilde{c}_{HB} \tilde{\cal O}_{HB}, 
\end{eqnarray*}
where $\bar{c}_i$ and $\tilde c_i$ are `Wilson coefficients' specifying the strength of the new CP-even and CP-odd interactions, respectively, and the dimension-six operators ${\cal O}_i$ are those described in Refs.~\cite{Contino:2013kra,Alloul:2013naa}. In the SM, all of the Wilson coefficients are equal to zero. The ${\cal O}_\gamma$ and $\tilde{\cal O}_\gamma$ operators introduce new interactions between the Higgs boson and two photons. The  ${\cal O}_g$ and $\tilde{\cal O}_g$ operators introduce new interactions between the Higgs boson and two gluons and the analysis presented in this Letter is sensitive to these operators  through the gluon fusion production mechanism. The ${\cal O}_{HW}$ and $\tilde{\cal O}_{HW}$ operators introduce new $HWW$, $HZZ$ and $HZ\gamma$ interactions. The  $HZZ$ and $HZ\gamma$ interactions are also impacted by ${\cal O}_{HB}$ and $\tilde{\cal O}_{HB}$ and, to a lesser extent, ${\cal O}_\gamma$ and $\tilde{\cal O}_\gamma$. The analysis presented in this Letter is sensitive to the ${\cal O}_{HW}$, $\tilde{\cal O}_{HW}$, ${\cal O}_{HB}$ and $\tilde{\cal O}_{HB}$ operators through vector-boson fusion and associated production.

Other operators in the full effective Lagrangian of Ref.~\cite{Contino:2013kra} can also modify Higgs boson interactions. 
Combinations of some of the CP-even operators have been constrained using global fits to experimental data from LEP and the LHC~\cite{Contino:2013kra,Pomarol:2013zra,Ellis:2014jta}.

\section{Statistical correlations between differential distributions}\label{sec:corr}

ATLAS~\cite{cite:atlas} is a multipurpose particle physics detector with cylindrical geometry and nearly $4\pi$ coverage in solid angle.\footnote{ATLAS uses a right-handed coordinate system with its origin
at the nominal interaction point (IP) at the centre of the detector and the $z$-axis along the beam
pipe. The $x$-axis points from the IP to the centre of the LHC ring, and the $y$-axis points
upward. Cylindrical coordinates $(r,\phi)$ are used in the transverse plane, $\phi$ being the
azimuthal angle around the beam pipe.} The analysis is performed using proton-proton collision data at a centre-of-mass energy $\sqrt{s}=8$~TeV and an integrated luminosity of 20.3~fb$^{-1}$.

\begin{figure}[t]
\centering
\includegraphics[width=\widthfactor\linewidth]{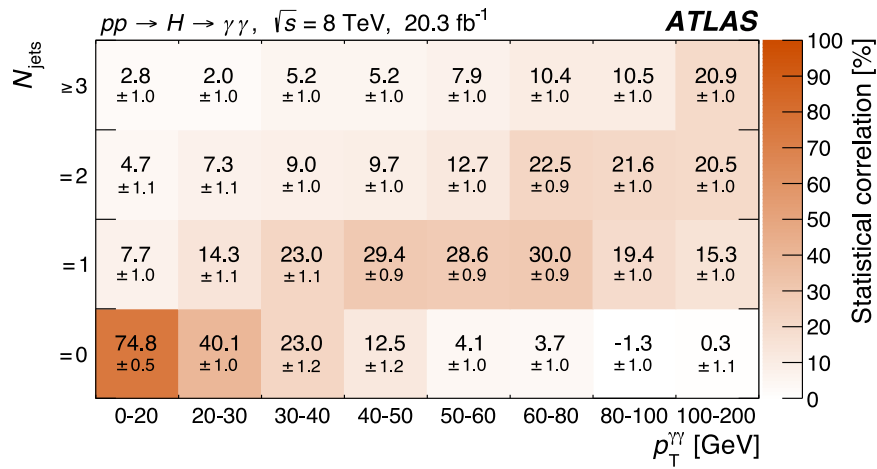}
\caption{\label{fig:minicorr}
Statistical correlations between the measured cross sections in bins of the diphoton transverse momentum and jet multiplicity distributions. The quoted uncertainties refer to the total statistical uncertainty due to the finite number of bootstrapped samples and the finite number of data events. 
}
\end{figure}

The object and event selections used to define the differential distributions are described in detail in Ref.~\cite{Aad:2014lwa}. 
The statistical correlations between the measured cross sections as a function of different distributions are obtained using a random sampling with replacement method on the detector-level data. This procedure is often referred to as `bootstrapping'~\cite{Hayes:1988xc}. Bootstrapped event samples are constructed from the data by assigning each event a weight pulled from a Poisson distribution with unit mean. The five differential distributions are then reconstructed using the weighted events, and the signal yields in each bin of a differential distribution are determined using an unbinned maximum-likelihood fit of the diphoton invariant mass spectrum (full details of the fit can be found in Ref.~\cite{Aad:2014lwa}).
The procedure is repeated 10000 times with statistically independent weights and the correlation between two bins of different distributions is determined from the scatter graph of the corresponding extracted cross sections. 
The observed correlations between bins of the measured $p_{\rm T}^{\gamma\gamma}$ and $N_{\rm jets}$ cross sections are shown in Fig.~\ref{fig:minicorr}.

The statistical uncertainties on the correlation due to the finite number of bootstrap samples ranges from 0.5\% to 1\%. The statistical uncertainty on the correlations due to the finite number of events in data is determined to be less than 2\% using the statistical overlap and variance of signal and background events in a mass window around the Higgs boson mass. In order to validate this approach, a set of pseudo-experiments was created from input conditions (with known correlations) chosen to be similar to those in data in terms of purity, kinematics and sample size. 
For each pseudo-experiment, a value for the correlation is determined using 10000 bootstrapped samples and compared to the input correlations. No bias due to the bootstrapping is observed in the central value obtained from 500 pseudo-experiments.    

As part of this Letter, the correlations computed above are made publicly available in HEPDATA~\cite{hepdata}, allowing the analysis to be repeated using alternative effective Lagrangians, complete EFT frameworks, or other models with non-SM Higgs boson interactions.

\section{Theoretical predictions}\label{sec:efthyspo}

The effective Lagrangian has been implemented in FeynRules \cite{Alloul:2013naa}.\footnote{The implementation in Ref.~\cite{Alloul:2013naa} involves a redefinition of the gauge boson propagators that results in unphysical amplitudes unless certain physical constants are also redefined. The original implementation did not include the redefinition of these physical constants. However, the impact of redefining the physical constants is found to be less than 1\% on the predicted cross sections across the range of Wilson coefficients studied. The relative change in the predicted Higgs boson cross sections as functions of the different Wilson coefficients is also found to agree with that predicted by the Higgs characterisation framework~\cite{Artoisenet:2013puc}, with less than 2\% variation across the parameter ranges studied.} 
Parton-level event samples are produced for specific values of Wilson coefficients by interfacing the universal file output from FeynRules to the \madgraph5~\cite{Alwall:2011uj} event generator. Higgs boson production via gluon fusion is produced with up to two additional partons in the final state using leading-order matrix elements. The 0-, 1- and 2-parton events are merged using the MLM matching scheme~\cite{Mangano:2006rw} and passed through the \pythia6 generator~\cite{Sjostrand:2006za} to create the fully hadronic final state. Event samples containing a Higgs boson produced either in association with a vector boson or via vector-boson fusion are produced using leading-order matrix elements and passed through the \pythia6 generator. For each production mode, the Higgs boson mass is set to 125~GeV~\cite{Aad:2015zhl} and events are generated using the CTEQ6L1 parton distribution function and the AUET2 parameter set~\cite{ATLASUE}. All other Higgs boson production modes are assumed to occur as predicted by the SM.

Event samples are produced for different values of a given Wilson coefficient.  The particle-level differential cross sections are produced using {\sc Rivet}~\cite{Buckley20132803}. 
The {\sc Professor} method~\cite{Buckley:2009bj} is used to interpolate between these samples, for each bin of each distribution, and provides a parameterisation of the effective Lagrangian prediction. 
The parameterisation function is determined using 11 samples when studying a single Wilson coefficient, whereas 25 samples are used when studying two Wilson coefficients simultaneously. 
As the Wilson coefficients enter the effective Lagrangian in a linear fashion, second-order polynomials are used to predict the cross sections in each bin. The method was validated by comparing the differential cross sections obtained with the parameterisation function to the predictions obtained with dedicated event samples generated at the specific point in parameter space. 

The model implemented in FeynRules fixes the Higgs boson width to be that of the SM, $\Gamma_{H} = 4.07$~MeV ~\cite{Heinemeyer:2013tqa}. The cross sections are scaled by $\Gamma_{H} / (\Gamma_{H} + \Delta \Gamma)$, where  $\Delta \Gamma$ is the change in partial width due to a specific choice of Wilson coefficient. The change in partial width is determined for each Higgs coupling using the partial-width calculator in \madgraph5 and normalised to reproduce the SM prediction from {\sc Hdecay} \cite{Djouadi:1997yw}. 

The leading-order predictions obtained from \madgraph5 are reweighted to account for higher-order QCD and electroweak corrections to the SM process, assuming that these corrections factorise from the new physics effects. The differential cross section as a function of variable $X$ for a specific choice of Wilson coefficient, $c_i$ is  given by
\begin{equation*} \label{eq:sigma}
\frac{{\rm d} \sigma}{{\rm d}X} = \sum_j \left( \frac{{\rm d}\sigma_j}{{\rm d}X} \right)^{\rm ref} \cdot \left( \frac{{\rm d}\sigma_j}{{\rm d}X} \right)_{c_i}^{\rm MG5} / \left( \frac{{\rm d}\sigma_j}{{\rm d}X} \right)_{c_i=0}^{\rm MG5},
\end{equation*}
where the summation $j$ is over the different Higgs boson production mechanisms, `MG5' labels the \madgraph5 prediction and `ref' labels a 
reference sample for SM Higgs boson production.

The reference sample for Higgs boson production via gluon fusion is simulated using MG5\_aMC@NLO~\cite{mg5aMCatNLO} with the CT10 parton distribution function~\cite{Lai:2010vv}. The $H+n$-jets topologies are generated using next-to-leading-order (NLO) matrix elements for each parton multiplicity ($n =$ 0, 1 or 2) and combined using the FxFx merging scheme \cite{Frederix:2012ps}. The parton-level events are passed through \pythia8~\cite{Sjostrand:2007gs} to produce the hadronic final state using the AU2 parameter set \cite{ATLASUE2}. The sample is normalised to the total cross section predicted by a next-to-next-to-leading-order plus next-to-next-to-leading-logarithm (NNLO+NNLL) QCD calculation with NLO electroweak corrections applied~\cite{Heinemeyer:2013tqa}. 
The reference sample for Higgs boson production via vector-boson fusion (VBF) is generated at NLO accuracy in QCD using the \powhegbox\ \cite{Nason:2009ai}. The events are generated using the CT10 parton distribution function (PDF) and \pythia8\ with the AU2 parameter set. The VBF sample is normalised to an approximate-NNLO QCD cross section with NLO electroweak corrections applied~\cite{Heinemeyer:2013tqa}. 
The reference samples for Higgs boson production in association with a vector boson ($VH$, $V$$=$$W$,$Z$) or a top--antitop pair ($t\bar{t}H$) are produced at leading-order accuracy using \pythia8 with the CTEQ6L1 PDF and the 4C parameter set~\cite{ATLASUE}. The $ZH$ and $WH$ samples are normalised to cross sections calculated at NNLO in QCD with NLO electroweak corrections, whereas the 
$t\bar{t}H$ sample is normalised to a cross section calculated to NLO in QCD~\cite{Heinemeyer:2013tqa}. 

\begin{figure}[t!]
\centering
\includegraphics[width=\widthfactor\linewidth]{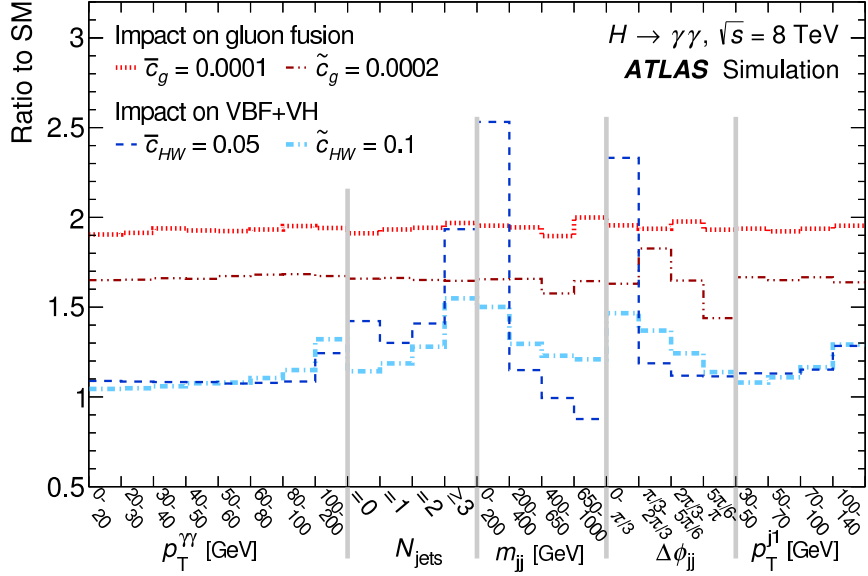}
\caption{
Ratio of differential cross sections predicted by specific choices of Wilson coefficient to the differential cross sections predicted by the SM.\label{fig:ratios}
}
\end{figure}

The ratio of the differential cross sections to the SM predictions for some representative values of the Wilson coefficients are shown in Fig.~\ref{fig:ratios}. The impact of the $\bar{c}_{g}$ and $\tilde{c}_{g}$ coefficients are presented for the gluon fusion production channel and show a large change in the overall cross section normalisation. The $\tilde{c}_{g}$ coefficient also changes the shape of the $\Delta\phi_{jj}$ distribution, which is expected from consideration of the tensor structure of CP-even and CP-odd interactions~\cite{Klamke:2007cu, Andersen:2010zx}. The impact of the $\bar{c}_{HW}$ and $\tilde{c}_{HW}$ coefficients are presented for the VBF$+$VH production channel and show large shape changes in all of the studied distributions.\footnote{Form factors are sometimes used to regularise the change of the cross section above a momentum scale $\Lambda_{\rm FF}$. This was investigated by reweighting the VBF$+$VH samples using form-factor predictions from {\sc VBFNLO~\cite{Arnold:2008rz}}. The impact on the $\bar c_{HW}$ and $\tilde{c}_{HW}$ limits are negligible for $\Lambda_{\rm FF}>$1~TeV.} The $\Delta\phi_{jj}$ distribution is known to discriminate between CP-odd and CP-even interactions in the VBF production channel~\cite{Plehn:2001nj}.

\section{Limit-setting procedure}\label{sec:limits}

Limits on the Wilson coefficients are set by constructing a $\chi^2$ function
\begin{align} \label{eq:chi2}
\chi^2 & = \left( \vec \sigma_{\text{data}} - \vec \sigma_{\text{pred}} \right)^{\rm T} \, C^{-1}\,  \left( \vec \sigma_{\text{data}} - \vec \sigma_{\text{pred}}\right) \, , \nonumber
\end{align}
where $\vec \sigma_{\rm data}$ and $\vec \sigma_{\rm pred}$ are vectors
from the measured and predicted cross sections of the five analysed observables, and
$C = C_{\rm stat} + C_{\rm exp} + C_{\rm pred}$ is the total covariance matrix defined by the sum of the statistical, experimental and theoretical covariances.
The predicted cross section $\vec \sigma_{\rm pred}$ and its associated covariance $C_{\rm pred}$ are
continuous functions of Wilson coefficients. Scans of one or two Wilson coefficients are carried out and the minimum $\chi^2$ value, $\chi^2_{\rm min}$, is determined. The confidence level ($CL$) of each scan point can be calculated as
\begin{align}
 1 - CL & = \,  n \, \int_{\chi^2(c_i) - \chi^2_{\rm min}}^{\infty} \text{d} x \, f( x; m) \, , \nonumber
\end{align}
with $\chi^2(c_i)$ being the $\chi^2$ value evaluated for a given Wilson coefficient $c_i$, and $f( x;m)$ being the $\chi^2$ distribution for $m$ degrees of freedoms and $n = 1$ or $\frac{1}{2}$ for two-sided or one-sided limits. 
The coverage of $CL$ and the effective number of degrees of freedom are determined using ensembles of pseudo-experiments.\footnote{For one-dimensional limits on the CP-even (odd) Wilson coefficients, good agreement is found between the asymptotic formula and the pseudo-experiment test statistic with $m=1$ and $n=1$ ($\frac{1}{2}$). For the two-dimensional limits on $\bar c_g$ versus $\tilde c_g$, and $\bar c_{HW}$ versus $\tilde c_{HW}$, good agreement between pseudo-experiments and asymptotic formula is found for $m = 1$ and $n=1$. For the two dimensional limit on $\bar c_g$ versus $\bar c_\gamma$, good agreement between pseudo-experiments and asymptotic formula is found for $m=2$ and $n =1$.}

\begin{figure}[t]
\centering
\includegraphics[width=\widthfactor\linewidth]{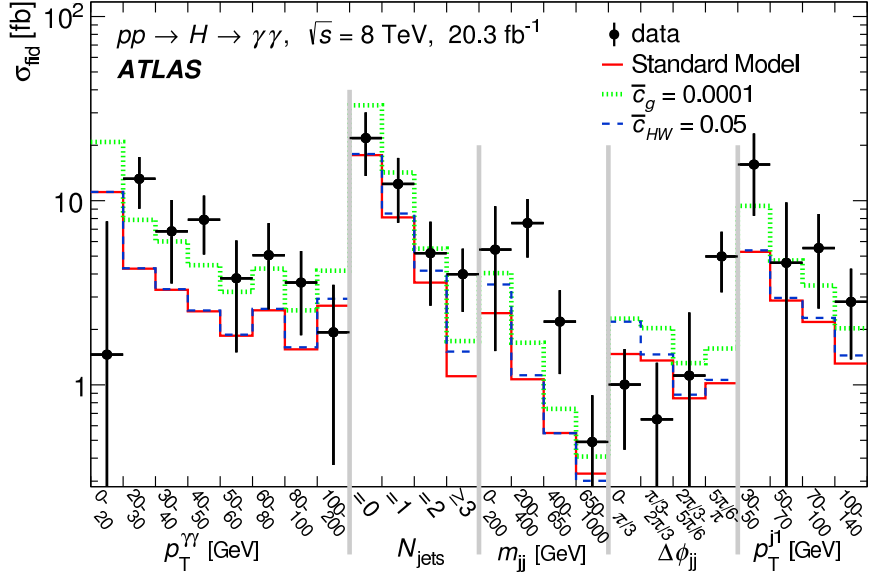}
\caption{The input data from Ref.~\cite{Aad:2014lwa} is compared to the SM hypothesis and two non-SM hypotheses with $\bar{c}_{g}=1\times10^{-4}$ and $\bar{c}_{HW}=0.05$, respectively.
\label{fig:super-plot}}
\end{figure}

The input data vector is compared in Fig.~\ref{fig:super-plot} to the SM hypothesis as well as two non-SM hypotheses specified by $\bar{c}_{g}=1\times10^{-4}$ and $\bar{c}_{HW}=0.05$, respectively.

The covariance matrix for experimental systematic uncertainties is constructed from all uncertainty sources provided by Ref.~\cite{Aad:2014lwa}, which include the jet energy scale and resolution uncertainties, photon energy and resolution uncertainties, and model uncertainties. Identical sources are assumed to be fully correlated across bins and variables and the sign of an error amplitude is taken into account when computing the covariance matrix. The statistical uncertainties on the cross correlation have a negligible impact on the results reported here.

The covariance matrix for the theoretical uncertainties  is constructed to account for missing higher-order corrections and PDF uncertainties in the SM reference predictions. The uncertainties in the gluon fusion reference samples are: (i) a shape uncertainty, estimated by simultaneously varying the factorisation and renormalisation scales in MG5\_aMC@NLO by a factor of 0.5 or 2.0, and (ii) uncertainties on the NNLO+NNLL QCD plus NLO electroweak total cross-section prediction~\cite{Heinemeyer:2013tqa}, arising from missing higher-order corrections and PDF uncertainties; these uncertainties are assumed to be fully correlated among bins and observables. For VBF, $ZH$ and $WH$, shape uncertainties are neglected because their impact is expected to be negligible with respect to all other theory uncertainties. Normalisation uncertainties for these processes are taken from Ref.~\cite{Heinemeyer:2013tqa}. 

The benefit of using more than one differential distribution in the analysis is quantified using an `Asimov dataset', which is a representative dataset of the median expected cross-section measurement  assuming the SM. 
For $\bar{c}_{g}$ and $\tilde{c}_{g}$, the use of a single inclusive distribution ($p_{\rm T}^{\gamma\gamma}$ or $N_{\rm jets}$) results in the same expected limits as the full five-dimensional fit. 
For $\bar{c}_{\gamma}$ and $\tilde{c}_{\gamma}$, the most sensitive variable is found to be $p_{\rm T}^{\gamma\gamma}$, with a 5\% improvement in the expected limits obtained from using the five-dimensional information.  
For $\bar{c}_{HW}$ and $\tilde{c}_{HW}$, the most sensitive variable is $\Delta\phi_{jj}$ and an 18\% improvement in the expected limits is obtained from using the five-dimensional fit. 
In summary, the expected sensitivity for $\bar{c}_{g}$, $\tilde{c}_{g}$, $\bar{c}_{\gamma}$ and $\tilde{c}_{\gamma}$ arises mainly from the normalisation of the different production mechanisms, and can be probed using the inclusive distributions that distinguish between the different processes, whereas the  $\bar{c}_{HW}$ and $\tilde{c}_{HW}$ coefficients benefit more from the full five-dimensional information due to the induced shape changes in the kinematics of the VBF$+$VH process.

\begin{figure}[t!]
\centering
\includegraphics[width=\widthfactor\linewidth]{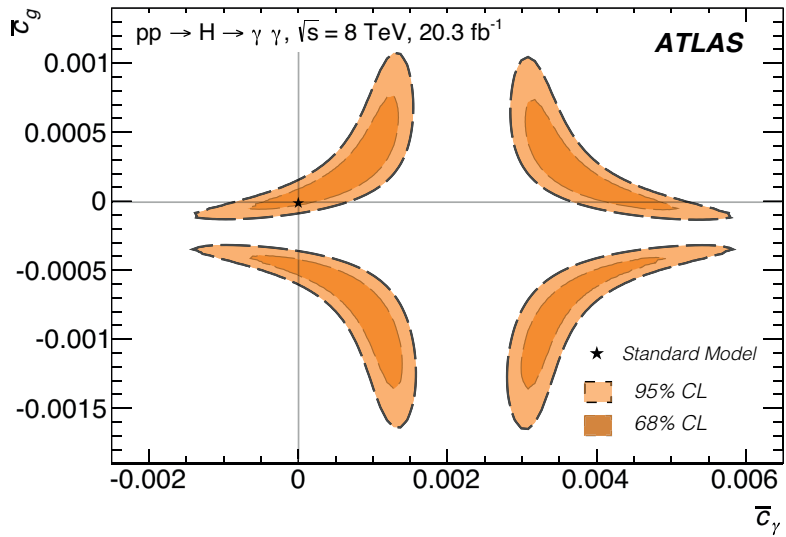}
\caption{
The 68\% (dark) and 95\% (light) confidence regions for the fit to the $\bar{c}_\gamma$ and $\bar{c}_g$ Wilson coefficients. All other coefficients are set to zero. The shaded area represents the allowed region of parameter space and the marker indicates the  SM value.
\label{fig:cy_cg}
}
\end{figure}

\section{Results}\label{sec:results}

The 68\% and 95\% confidence regions for a two-dimensional scan of $\bar c_{\gamma}$ and $\bar c_{g}$ are shown in Fig. \ref{fig:cy_cg}, after setting all other Wilson coefficients to zero. 
These additional interactions can interfere with the corresponding SM interactions. Destructive interference, for example, causes the $H \rightarrow \gamma \gamma$ branching ratio to be zero at $\bar c_\gamma \sim 2\times10^{-3}$ and the gluon fusion production cross section to be zero at $\bar c_g \sim -2.2\times10^{-4}$. The impact of these effects is evident in the structure of the obtained limits in the two-dimensional parameter plane. 

\begin{figure}[t]
\centering
\includegraphics[width=\widthfactor\linewidth]{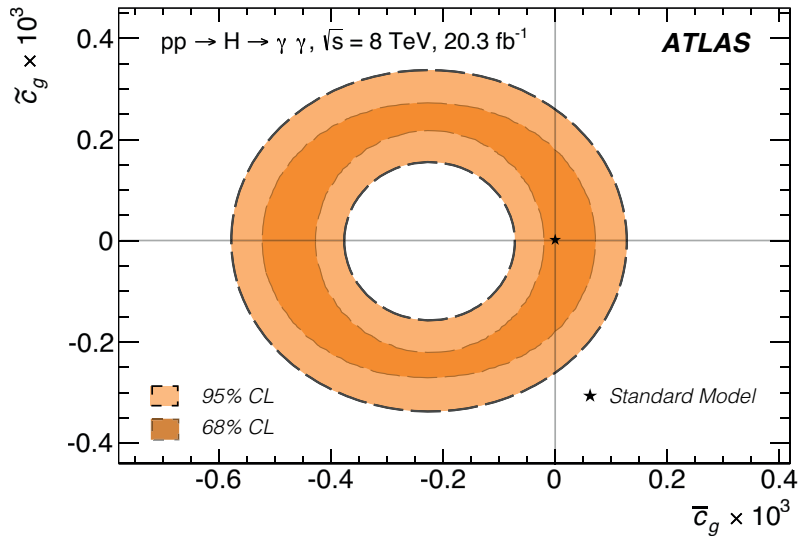}
\caption{
The 68\% (dark) and 95\% (light) confidence regions for the fit to the $\bar{c}_g$ and $\tilde{c}_g$ Wilson coefficients. All other  coefficients are set to zero. The shaded area represents the allowed region of parameter space and the marker indicates the SM value.  \label{fig:cg_cg}
}
\end{figure}

The 68\% and 95\% confidence regions for a two-dimensional scan of $\bar c_{g}$ and $\tilde{c}_{g}$ are shown in Fig. \ref{fig:cg_cg}, after setting all other Wilson coefficients to zero. 
The $\Delta\phi_{jj}$ distribution is sensitive to the $\tilde{c}_{g}$ parameter through the gluon fusion production mechanism (Figs.~\ref{fig:ratios} and \ref{fig:super-plot}) and the limit on $\tilde{c}_{g}$ is improved with the inclusion of this data in the fit. This is evident in Fig. \ref{fig:cg_cg} where the limit band is constricted at the largest values of $\tilde{c}_{g}$.

\begin{figure}[t]
\centering
\includegraphics[width=\widthfactor\linewidth]{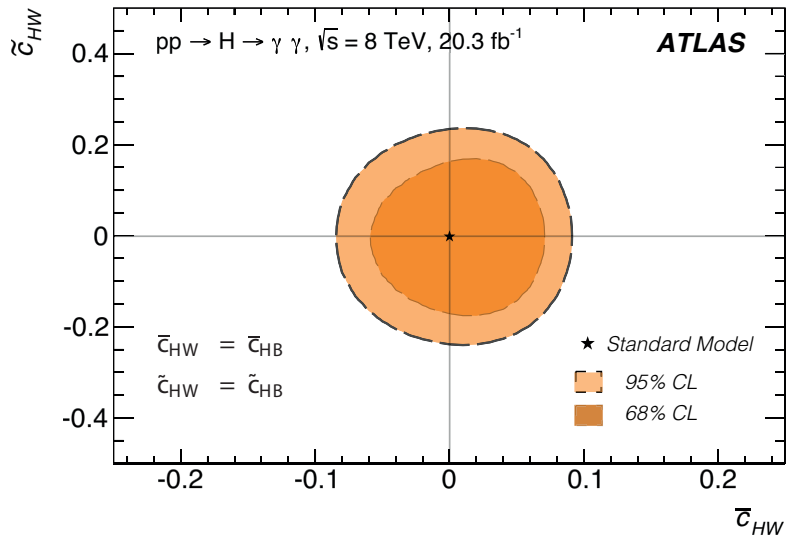}
\caption{
The 68\% (dark) and 95\% (light) confidence regions for the fit to the $\bar c_{HW}$ and $\tilde c_{HW}$ Wilson coefficients. 
All other Wilson coefficients are set to zero, except for $\bar c_{HB}$ and $\tilde c_{HB}$ which are set to be equal to $\bar c_{HW}$ and $\tilde c_{HW}$, respectively. 
The shaded area represents the allowed region of parameter space and the marker indicates the SM value. \label{fig:cHW_cHB}}
\end{figure}

The 68\% and 95\% confidence regions obtained from scanning $\bar c_{HW}$ and $\tilde{c}_{HW}$ are shown in Fig. \ref{fig:cHW_cHB}, after setting $\bar c_{HB}=\bar c_{HW}$ and $\tilde{c}_{HB}=\tilde{c}_{HW}$ to ensure that the partial width for $H \rightarrow Z \gamma$ is unchanged from the SM prediction.\footnote{Values of $\left| \bar c_{HW} - \bar c_{HB} \right| > 0.033$ lead to a very large decay rate for the $H\rightarrow Z\gamma$ process that is contradicted by the experimental constraints reported by ATLAS~\cite{Aad:2014fia}.} 
 As discussed in Section~\ref{sec:limits}, these Wilson coefficients produce large shape changes in all distributions and the obtained limits are strongest when fitting all five distributions simultaneously. 

The 95\% confidence regions for $\bar c_{HW}$ and $\tilde{c}_{HW}$ can be translated into the Higgs Characterisation framework~\cite{Artoisenet:2013puc} and compared to the ATLAS results for non-SM CP-even and CP-odd $HVV$ interactions, which were obtained using an angular analysis of the decay products in the $WW^*$ and $ZZ^*$ decay channels \cite{Aad:2015mxa}. The translated limits are $-0.08 < \tilde \kappa_{\rm HVV} / \kappa_{\rm SM} < 0.09 $ and $-0.22 < {\rm tan}(\alpha) \cdot \tilde \kappa_{\rm AVV} / \kappa_{\rm SM} < 0.22$, where the variables $\tilde \kappa_{\rm HVV}$, $\tilde \kappa_{\rm AVV}$, $ \kappa_{\rm SM}$ and $\alpha$ are defined in Refs.~\cite{Artoisenet:2013puc,Aad:2015mxa}. The limits obtained in this analysis are a factor of approximately seven stronger than those in Ref.~\cite{Aad:2015mxa}, due to increased sensitivity to the different Higgs boson production channels arising from the inclusion of rate and jet kinematic information in the signal hypothesis.

The observed limits on $\bar c_{HW}$ and $\tilde{c}_{HW}$ are also not excluded by current signal strength measurements. For example, the signal strength in the $H \rightarrow ZZ^*$ and $H\rightarrow WW^*$ channels is predicted to be approximately 1.3 for $\bar c_{HW}=0.1$, which is consistent with the dedicated measurements~\cite{ATLAS:2014aga,Aad:2014eva}. 

The 95\% confidence regions for a one-dimensional scan of the Wilson coefficients are given in Table~\ref{tab:1D_scans}. 

\begin{table}[t]
\centering

\vspace{2ex}
\begin{tabular}{| l |  l |}
\hline
 Coefficient & 95\% $1-CL$ limit  \\
\hline
$\bar c_{\gamma}$ & \cyreslimone\ $\cup$ \cyreslimtwo \\ 
$\tilde c_{\gamma}$ & \tcyreslim  \\
$\bar c_{g}$  & \cgreslimtwo\ $\cup$ \cgreslimone\ \\
$\tilde c_{g}$ & \tcgreslim  \\
$\bar c_{HW}$ & \cHWreslim \\
$\tilde c_{HW}$ & \tcHWreslim \\
\hline
\end{tabular}
\caption{ Observed allowed ranges at 95\% CL for the $\bar c_\gamma$, $\bar c_g$ and $\bar c_{HW}$ Wilson coefficients and their CP-conjugate partners. Limits on $\bar c_\gamma$, $\bar c_g$, $\tilde c_\gamma$ and $\tilde c_g$ are each derived with all other Wilson coefficients set to zero. Limits on $\bar c_{HW}$ and $\tilde c_{HW}$ are derived with $\bar c_{HB} = \bar c_{HW}$ and  $\tilde c_{HB} = \tilde c_{HW}$, respectively. Two allowed regions are observed for $\bar c_\gamma$ and $\bar c_g$, with the region between the solutions producing too small $pp\rightarrow H \rightarrow \gamma \gamma$ cross section due to destructive interference between new interactions and the SM. \label{tab:1D_scans} }
\end{table}

\section{Summary}

The strength and structure of the Higgs boson's interactions with other particles have been investigated using an effective Lagrangian. 
Limits are placed on anomalous CP-even and CP-odd interactions between the Higgs boson and photons, gluons, $W$-bosons and $Z$-bosons, using a fit to five differential cross sections previously measured by ATLAS in the $H \to \gam\gam$ decay channel at $\sqrt{s}=8$~TeV \cite{Aad:2014lwa}. No significant deviations from the SM predictions are observed. To allow a simultaneous fit to all distributions, the statistical correlations between these distributions have been determined by re-analysing the candidate $H \to \gam\gam$ events in the proton-proton collision data. These correlations are made publicly~\cite{hepdata} available to allow for future analysis of theories with non-SM Higgs boson interactions.

\section*{Acknowledgements}
% Acknowledgements for papers with collision data
% Version 15-Sep-2015

% Standard acknowledgements start here
%----------------------------------------------
We thank CERN for the very successful operation of the LHC, as well as the
support staff from our institutions without whom ATLAS could not be
operated efficiently. We also thank B.~Fuks and V.~Sanz for 
clarifications and calculations regarding the effective Lagrangian implementation
used in this article.

We acknowledge the support of ANPCyT, Argentina; YerPhI, Armenia; ARC, Australia; BMWFW and FWF, Austria; ANAS, Azerbaijan; SSTC, Belarus; CNPq and FAPESP, Brazil; NSERC, NRC and CFI, Canada; CERN; CONICYT, Chile; CAS, MOST and NSFC, China; COLCIENCIAS, Colombia; MSMT CR, MPO CR and VSC CR, Czech Republic; DNRF, DNSRC and Lundbeck Foundation, Denmark; IN2P3-CNRS, CEA-DSM/IRFU, France; GNSF, Georgia; BMBF, HGF, and MPG, Germany; GSRT, Greece; RGC, Hong Kong SAR, China; ISF, I-CORE and Benoziyo Center, Israel; INFN, Italy; MEXT and JSPS, Japan; CNRST, Morocco; FOM and NWO, Netherlands; RCN, Norway; MNiSW and NCN, Poland; FCT, Portugal; MNE/IFA, Romania; MES of Russia and NRC KI, Russian Federation; JINR; MESTD, Serbia; MSSR, Slovakia; ARRS and MIZ\v{S}, Slovenia; DST/NRF, South Africa; MINECO, Spain; SRC and Wallenberg Foundation, Sweden; SERI, SNSF and Cantons of Bern and Geneva, Switzerland; MOST, Taiwan; TAEK, Turkey; STFC, United Kingdom; DOE and NSF, United States of America. In addition, individual groups and members have received support from BCKDF, the Canada Council, CANARIE, CRC, Compute Canada, FQRNT, and the Ontario Innovation Trust, Canada; EPLANET, ERC, FP7, Horizon 2020 and Marie Skłodowska-Curie Actions, European Union; Investissements d'Avenir Labex and Idex, ANR, Region Auvergne and Fondation Partager le Savoir, France; DFG and AvH Foundation, Germany; Herakleitos, Thales and Aristeia programmes co-financed by EU-ESF and the Greek NSRF; BSF, GIF and Minerva, Israel; BRF, Norway; the Royal Society and Leverhulme Trust, United Kingdom.

The crucial computing support from all WLCG partners is acknowledged
gratefully, in particular from CERN and the ATLAS Tier-1 facilities at
TRIUMF (Canada), NDGF (Denmark, Norway, Sweden), CC-IN2P3 (France),
KIT/GridKA (Germany), INFN-CNAF (Italy), NL-T1 (Netherlands), PIC (Spain),
ASGC (Taiwan), RAL (UK) and BNL (USA) and in the Tier-2 facilities
worldwide.
%----------------------------------------------

\bibliographystyle{atlasBibStyleWoTitle}
\bibliography{EFTPaper}

\newpage
% ATLAS Collaboration author list
% Data extracted on 11-Aug-2015 for paper reference HIGG-2015-02
\begin{flushleft}
{\Large The ATLAS Collaboration}

\bigskip

G.~Aad$^{\rm 85}$,
B.~Abbott$^{\rm 113}$,
J.~Abdallah$^{\rm 151}$,
O.~Abdinov$^{\rm 11}$,
R.~Aben$^{\rm 107}$,
M.~Abolins$^{\rm 90}$,
O.S.~AbouZeid$^{\rm 158}$,
H.~Abramowicz$^{\rm 153}$,
H.~Abreu$^{\rm 152}$,
R.~Abreu$^{\rm 116}$
% ...
\bigskip
\\
% ...
$^{11}$ Institute of Physics, Azerbaijan Academy of Sciences, Baku, Azerbaijan\\
% ...
$^{85}$ CPPM, Aix-Marseille Universit{\'e} and CNRS/IN2P3, Marseille, France\\
% ...
$^{90}$ Department of Physics and Astronomy, Michigan State University, East Lansing MI, United States of America\\
% ...
$^{107}$ Nikhef National Institute for Subatomic Physics and University of Amsterdam, Amsterdam, Netherlands\\
% ...
$^{113}$ Homer L. Dodge Department of Physics and Astronomy, University of Oklahoma, Norman OK, United States of America\\
% ...
$^{116}$ Center for High Energy Physics, University of Oregon, Eugene OR, United States of America\\
% ...
$^{151}$ Institute of Physics, Academia Sinica, Taipei, Taiwan\\
$^{152}$ Department of Physics, Technion: Israel Institute of Technology, Haifa, Israel\\
$^{153}$ Raymond and Beverly Sackler School of Physics and Astronomy, Tel Aviv University, Tel Aviv, Israel\\
% ...
$^{158}$ Department of Physics, University of Toronto, Toronto ON, Canada\\
% ...
\end{flushleft}
% Created with ./xml2latex.py

\end{document}